\documentclass[acmsmall,nonacm]{acmart}

\makeatletter
\def\@ACM@checkaffil{
	\if@ACM@instpresent\else
	\ClassWarningNoLine{\@classname}{No institution present for an affiliation}%
	\fi
	\if@ACM@citypresent\else
	\ClassWarningNoLine{\@classname}{No city present for an affiliation}%
	\fi
	\if@ACM@countrypresent\else
	\ClassWarningNoLine{\@classname}{No country present for an affiliation}%
	\fi
}
\makeatother

\AtBeginDocument{%
  }

\usepackage{todonotes}
\usepackage{cleveref}
\usepackage{subfig}
\usepackage[utf8]{inputenc}
\usepackage{adjustbox}
\usepackage{threeparttable}
\usepackage{wrapfig}

\renewcommand\footnotetextcopyrightpermission[1]{} 

\newenvironment{bio}[1]
{\par
	\bigskip
	\begin{wrapfigure}{l}[0pt]{1in}
		\vspace{-15pt}
		\includegraphics[width=1in,height=1in,clip,keepaspectratio]{#1}
		\vspace{-5pt}
	\end{wrapfigure}
	\footnotesize \noindent}
{\par\bigskip}

\begin{document}
	
\pagenumbering{gobble}

\title{Are Software Updates Useless Against Advanced Persistent Threats?}
\author{Fabio Massacci}
\affiliation{%
	\institution{University of Trento and Vrije Universiteit Amsterdam}
}
\email{fabio.massacci@ieee.org}
\author{Giorgio Di Tizio}
\affiliation{%
	\institution{University of Trento}
}
\email{giorgio.ditizio@unitn.it}

\begin{figure}
	\centering
	\includegraphics[width=.3\textwidth]{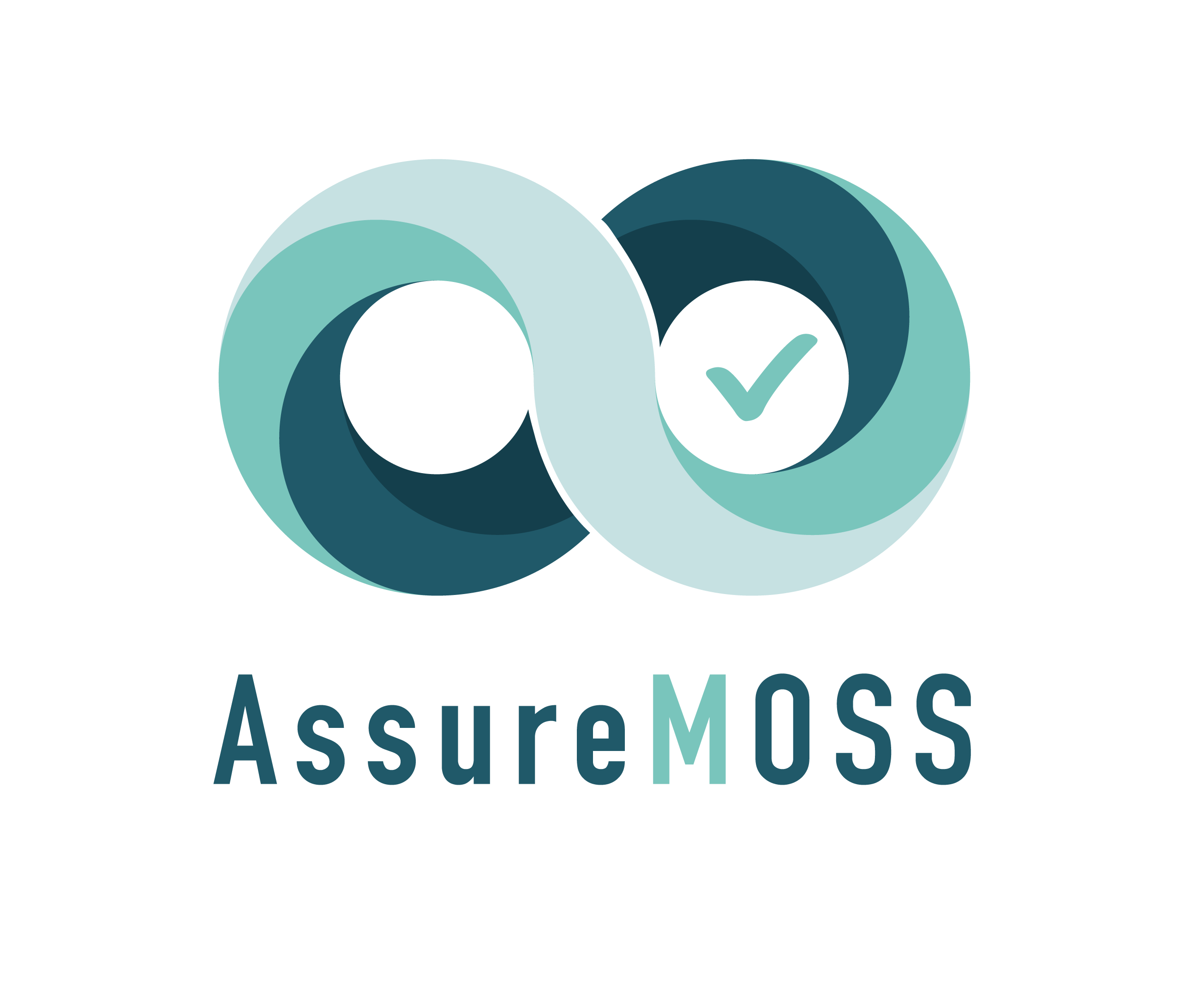}
\end{figure}

\vspace{2\baselineskip}

\vspace{2\baselineskip}

\begin{center}
	{\huge \textbf{Are Software Updates Useless Against Advanced Persistent Threats?}}
\end{center}

\vspace{\baselineskip}

{\large
	Authors:
	\begin{itemize}
		\item[] \textbf{Fabio Massacci}, University of Trento (IT), Vrije Universiteit Amsterdam (NL)    
		\item[] \textbf{Giorgio Di Tizio}, University of Trento (IT)
	\end{itemize}
}

\vfill

\begin{wrapfigure}{l}{2.5cm}
	\vspace{-\baselineskip}
	\includegraphics[width=2.5cm]{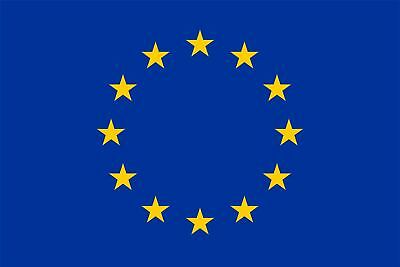}
\end{wrapfigure}

This paper was written within the H2020 AssureMOSS project that received funding 
from the European Union's Horizon 2020 research and innovation programme under 
grant agreement No 952647. This paper reflects only the author's view and the 
Commission is not responsible for any use that may be made of the information 
contained therein.

\clearpage
\twocolumn

\begin{bio}{images/Logo_colori}
	\textbf{Assurance and certification in secure Multi-party Open Software and 
		Services (AssureMOSS)} No single company does master its own national, in-house 
	software. Software is mostly assembled from “the internet” and more than half 
	come from Open Source Software repositories (some in Europe, most elsewhere). 
	Security \& privacy assurance, verification and certification techniques 
	designed for large, slow and controlled updates, must now cope with small, 
	continuous changes in weeks, happening in sub-components and decided by third 
	party developers one did not even know they existed. AssureMOSS proposes to 
	switch from process-based to artefact-based security evaluation by supporting 
	all phases of the continuous software lifecycle (Design, Develop, Deploy, 
	Evaluate and back) and their artefacts (Models, Source code, Container images, 
	Services). The key idea is to support mechanisms for lightweigth and scalable 
	screenings applicable automatically to the entire population of software 
	components by Machine intelligent identification of security issues, Sound 
	analysis and verification of changes, Business insight by risk analysis and 
	security evaluation. This approach supports fast-paced development of better 
	software by a new notion: continuous (re)certification. The project will 
	generate also benchmark datasets with thousands of vulnerabilities. AssureMOSS: 
	\textbf{Open Source Software: Designed Everywhere, Secured in Europe}. More 
	information at https://assuremoss.eu.
\end{bio}

\begin{bio}{photos/massacci}
	\textbf{Fabio Massacci} (PhD 1997) 	is a professor at the
	University of Trento, Italy, and Vrije Universiteit
	Amsterdam, The Netherlands. He received the Ten
	Years Most Influential Paper award by the IEEE
	Requirements Engineering Conference in 2015. He
	is the European Coordinator of the AssureMOSS
	project. Contact him at \emph{fabio.massacci@ieee.org}.
\end{bio}

\begin{bio}{photos/giorgio}
	\textbf{Giorgio Di Tizio} (PhD 2023) is a Postdoctoral researcher at the 
	University of Trento, Italy. His interests include cyber threat intelligence and cybercrime. Contact him at \emph{giorgio.ditizio@unitn.it}.
\end{bio}

How to cite this paper:
\begin{itemize}
	\item Massacci, F. and Di Tizio, G. Are Software Updates Useless Against Advanced Persistent Threats?. \emph{Communications of the ACM 66, 1 (2023)}. DOI: https://doi.org/10.1145/3571452
\end{itemize}

License:
\begin{itemize}
	\item This article is made available with a perpetual, non-exclusive, non-commercial license to distribute.
\end{itemize}

\clearpage

\pagenumbering{arabic}

\onecolumn
\maketitle
\thispagestyle{plain}
\pagestyle{plain}

A dilemma worth Shakespeare's Hamlet is increasingly haunting 
companies and security researchers: ``to update or not to 
update, this is the question``. From the perspective of 
recommended common practices by software vendors 
the answer is unambiguous: you should keep your software
up-to-date~\cite{DBLP:journals/ieeesp/ReederIC17}.
But is common sense always good sense? We argue it is not.

Last year on a CACM's column 
Poul-Henning Kamp~\cite{kamp2021software} argued that 
these industry best practices do not seem to work and a 
more radical reform is needed. In the same year,
on the IEEE S\&P magazine Massacci, Trent and Peisert 
reminds us that the SolarWinds attack was funneled by an 
update \cite{massacci2021solarwinds} and a second choral piece
this year \cite{massacci-etal-2022-protestware} tells us that the recent protestware attacks are 
also channeled through updates. 

What is badly wrong here is that updates are hardly classified as either functionality or security updates or both. They are bundled together for the convenience of the software vendor~\cite{massacci2021solarwinds}. For example, the WhatsApp update v2.19.51, while patching a critical security vulnerability exploited by the NSO Group, summarized the update with the following note: "You can now see stickers in full size when you long press a notification".
One might concede, without believing it, that conflating together functionality and security updates is done to make it more difficult to identify the vulnerable code.

Yet, this lack of transparency is not going to help. Organizations can only blindly accept the "black-box" cumulative update that will force them to install all updates ignored so far, or equally blindly ignore the popup. As Trent Jager said: damned if 
you patch, damned if you do not.

Still, updates 
might be
normally good and
might turn to be unwise only in the high-profile cases that hit the media.

We investigated whether this is the case in the context 
of Advanced Persistent Threats 
(APTs)~\cite{di2022software}, 'la creme de 
la creme' of the attacker ecosystem. APTs 
are sophisticated actors that deliberately 
and persistently target specific 
individuals and companies with a strategic 
motivation (from sabotage to financial 
gain). In this scenario, only an ‘all hands 
on deck’ defense seems appropriate
and keeping your software up-to-date 
seems the bare, and likely not even 
sufficient, minimum. 

Starting from 'Operation Aurora' the 
security community increasingly released 
public information about APTs campaigns via 
blogs and technical reports. Unfortunately, 
the information is fragmented over 
different sources, each using different 
taxonomies to track adversaries. 
So, we collected data about more than 350 APT campaigns performed by 86 APTs in more than 10 years from more than 500 resources (and, by the way, the data is open source\footnote{https://doi.org/10.5281/zenodo.6514817}).

From this wealth of data, we can try to understand if these Threats called APTs really deserve the acronym they got.

\begin{figure}
	\centering
	\captionsetup[subfigure]{labelformat=empty}
	\subfloat[More than half of the campaigns do not exploit software vulnerabilities but extensively rely on spearphishing to get initial access.]{
		\includegraphics[width=0.9\columnwidth,height=0.6\columnwidth]{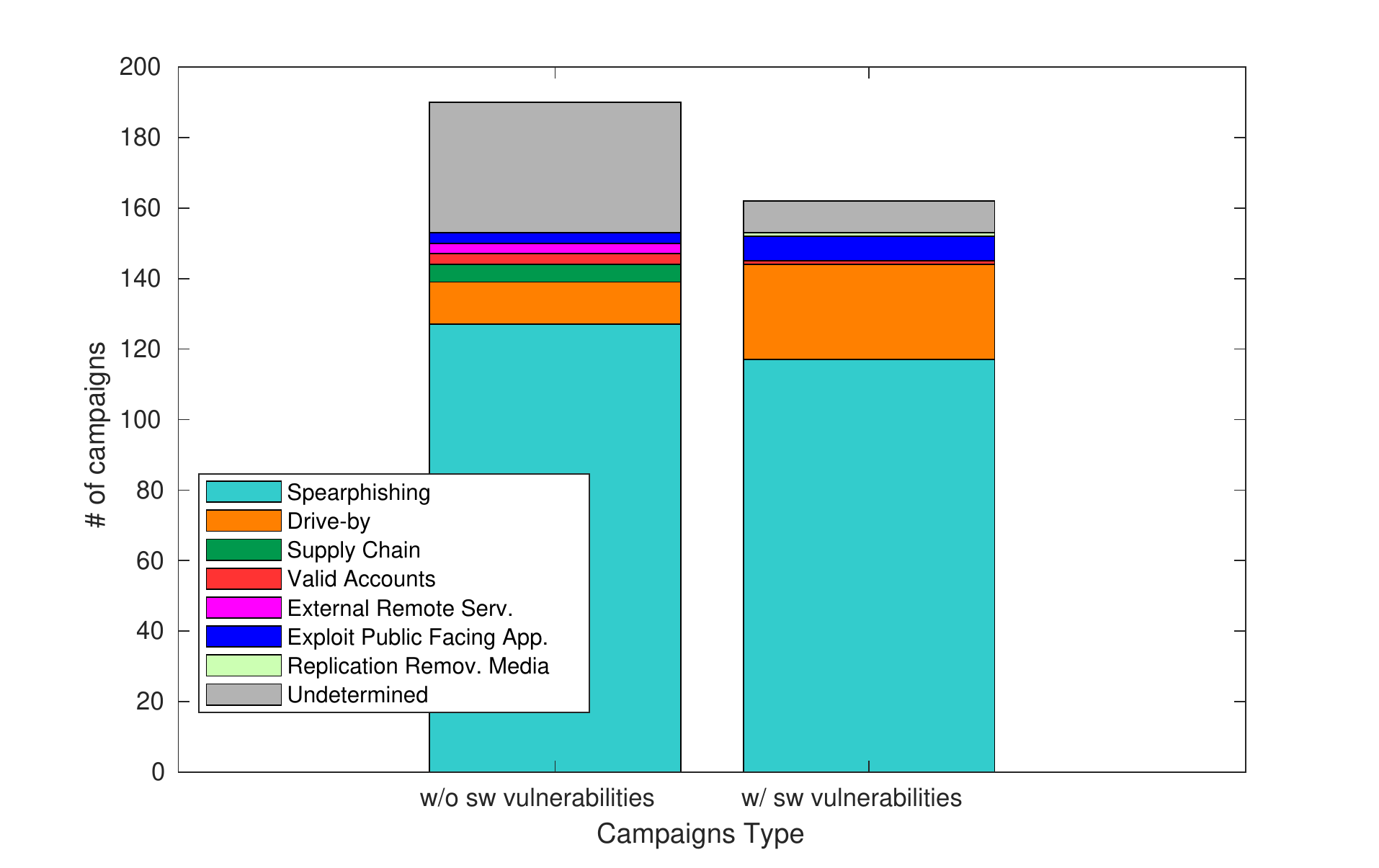}}
	\caption{Attack vector campaigns and software vulnerabilities}
\label{fig:attack_vectors}
\end{figure}

\paragraph*{A as Advanced:} In most cases, APTs do not even exploit a software vulnerability. \Cref{fig:attack_vectors} shows the attack vectors employed in the campaigns. More than half of them do not employ any software vulnerability. APTs rely on spearphishing attacks via email and social networks to obtain the initial footprint in the network. 

Whether your software is up to date or not makes no difference in at least 50\% 
of the cases, as software vulnerabilities are not the main attack vector. The 'all-powerful' 
attacker supporters 
could argue that we are a victim of 
survival bias as we used data of security attacks that have been eventually discovered.
As Richard Clayton said 
at the Workshop on Economics of Information 
Security on 2017, ``we don’t always know if are measuring
dim attackers getting caught rather than
smart attackers getting through``. This is a 
valid point if we consider a single point 
in time. Eventually, even smart attackers will be,
if not caught, at least spotted. For example, we can hardly say that the 
Stuxnet and Solarwinds attacks were 
performed by unskilled 
attackers. Still, they have 
eventually been in the news. Thus, a decade of reports can give a reasonably accurate view of 
the attacker ecosystem. 

So let's zoom in on the half of the attacks where software vulnerabilities did play some role.

\begin{figure}
\centering
	\captionsetup[subfigure]{labelformat=empty}
	\subfloat[Most of the campaigns exploited at least one vulnerability after reservation and publication by NVD. Only a few launched attacks exploiting unknown vulnerabilities that were neither reserved nor published by NVD.]{
		\includegraphics[width=0.9\columnwidth,height=0.6\columnwidth]{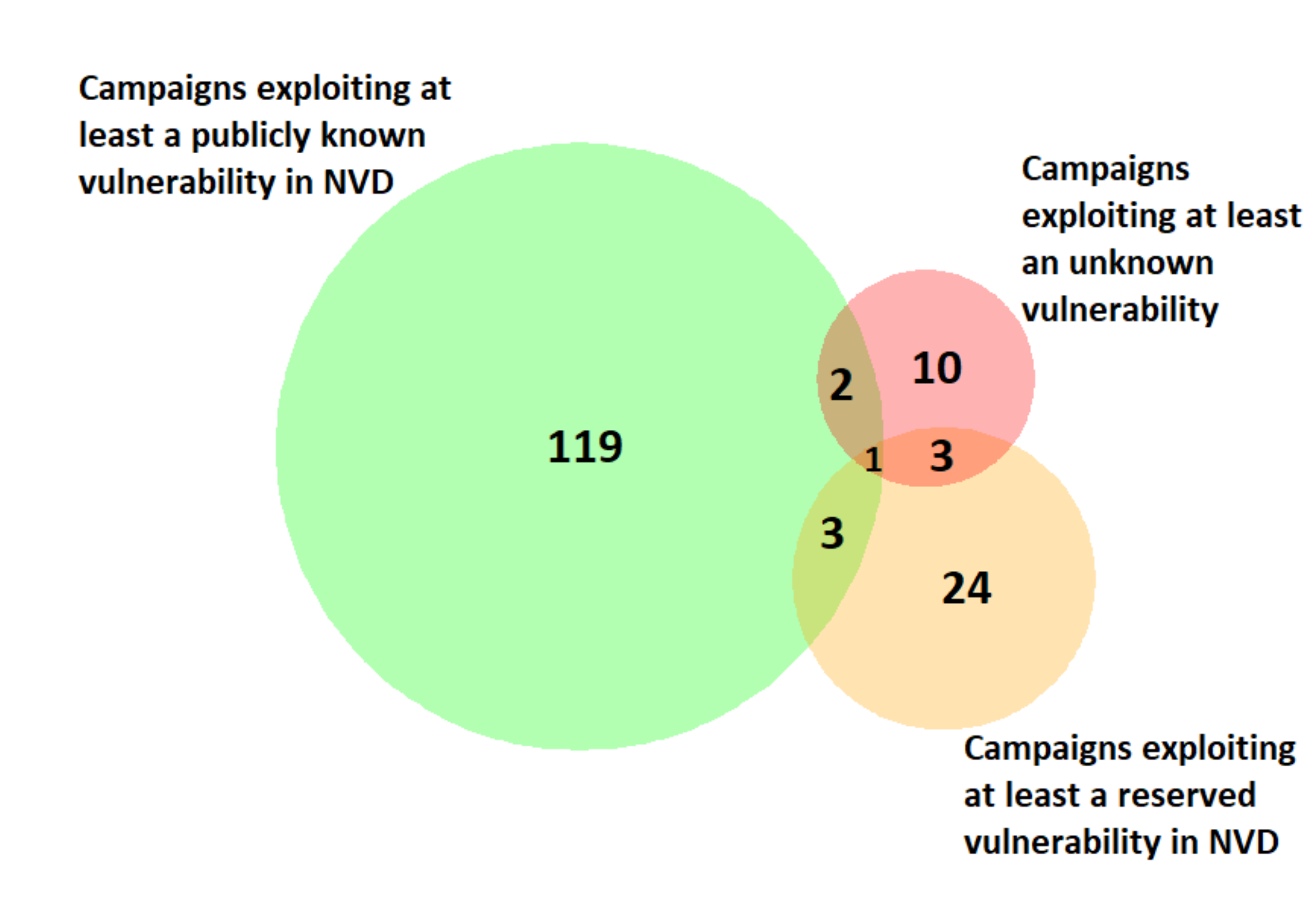}}
	\caption{Classification of APT Campaigns.}
	\label{fig:Venn-MITRE-NVD}
\end{figure}

\paragraph*{P as Persistent} When APTs exploit software vulnerabilities, they often exploit vulnerabilities of which the vendor is aware and for which a patch is available. The 0-days exploits are not the norm. \Cref{fig:Venn-MITRE-NVD} summarizes the number of campaigns that exploited known or 0-day software vulnerabilities. The 77\% of the campaigns in our dataset rely on at least a vulnerability that was previously published on NVD at the time of the attack. 

This finding seems pretty solid evidence that
software updates are actually useful. If APTs are exploiting software vulnerabilities that are known, and for which an update exists, then staying always on the edge of the newest Microsoft Office or Adobe Acrobat version should be enough to prevent the exploitation. Unfortunately, raw uninterpreted statistics of hindsight can be misleading.

The first problem with "Keep your 
software updated to the newest version" is 
that the pace of updates is often very 
close to a Denial of Service against IT
administrators with no improvement in 
security~\cite{DBLP:journals/tissec/AllodiM14}. Many updates are \emph{not} 
security updates and updates tend to break 
other functionalities. Huang et al. \cite{huang-updatability} in a paper aptly 
entitle up-to-crash found
out that many seemingly innocuous updates 
of vulnerable Android libraries might crash 
the dependent application with up to 50\% 
chances. Toss a coin, if head is up, software 
is down.

So we decided to investigate the potential 
effectiveness of keeping the software up-to-date for 5 widely used software products in 
the decade under study
(Office, Acrobat Reader, Air, JRE, and 
Flash Player) from three major software 
companies (Adobe, Microsoft, and Oracle) 
for the Windows O.S. with different 
strategies.  

The \emph{Ideal} strategy 
represents the unrealistic scenario
of staying on 
the edge of the wave and update the 
software as soon as a new release is 
available. The software vendor's speed is the only limiting factor. 

The \emph{Industry} strategy is a more realistic representation of the dutiful organizations that follow the industry best practices and apply each new update with some delay between the release of an update and its installation to perform regression testing. Delay ranges from one week to several months.

A more relaxed \emph{Reactive} strategy just focuses on 
updating when a CVE for a software 
vulnerability is published in a database of 
vulnerabilities like NVD. 

\paragraph*{T as Threat} As we (now) know when the attack took place 
with some approximation (i.e almost all reports only indicate the month of the attack), we can calculate the Agresti-Coull 95\% confidence interval that doing an update \emph{at that time} was actually useful or not. In other words, we can calculate the conditional probability that you would have updated on the same month, or updated the next month when the update  became available and still have succumbed to
an APT - if they ever decided to target you.

\begin{figure}
	\captionsetup[subfloat]{labelformat=empty}
	\centering
	\subfloat[In the best-case scenario, updates are deployed \emph{before} exploitation if they overlap on the same date. In this case, either an enterprise updates immediately or the risk of being compromised by updating only when a CVE is published is the same as always updating to the newest version. In the worst-case scenario, updates are deployed \emph{after} exploitation if they overlap on the same date. In this case, even if an enterprise updates immediately there is considerable overlap in the probability of being compromised with more relaxed approaches]{
		\includegraphics[width=0.9\columnwidth]{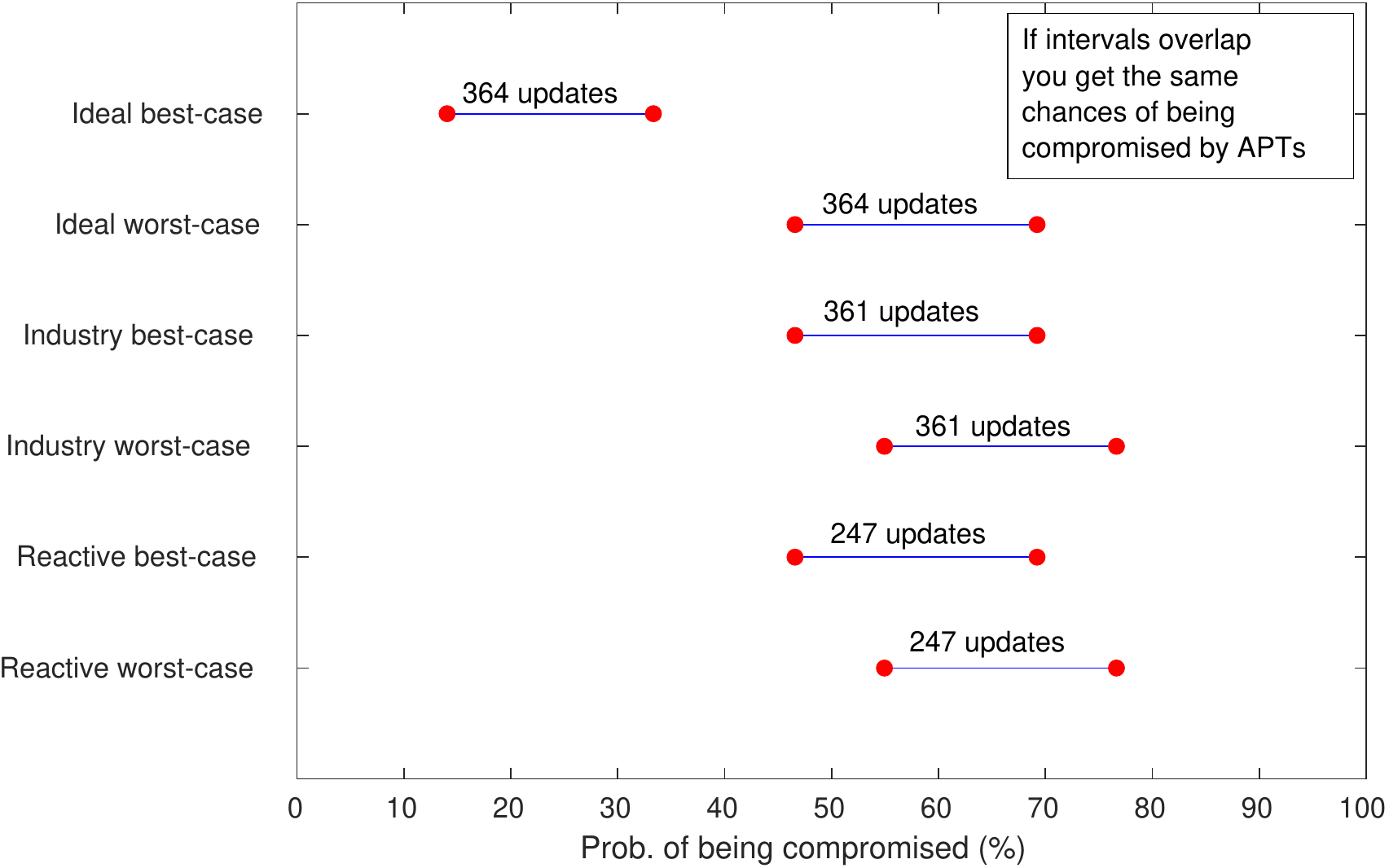}}
	\caption{Agresti-Coull confidence interval of the probability of being compromised for update strategies (one month delay for \emph{Industry} and \emph{Reactive} strategies)}
	\label{fig:agresti-intervals}
\end{figure}

\Cref{fig:agresti-intervals} shows the range in which the probability of
being compromised lies for the different strategies considering a one-month delay for the \emph{Industry} and \emph{Reactive} strategies to perform regression testing. When the intervals overlap it means the two strategies are not statistically distinguishable. 
That is a vey bad news for the IT administrators that struggled to keep pace of the updates.

In \Cref{fig:agresti-intervals} we observe that either an enterprise applies at light speed lots of updates (the \emph{Ideal} strategy) or the risk of being compromised once you have to wait to perform regression testing for all new releases of the software (\emph{Industry} strategy) is the same as if you just wait for a CVE to come out (\emph{Reactive} strategy) but the latter saves you from a lot of useless updates. Furthermore, even if we are keeping our system up to date at light speed, we also need to be lucky to not jeopardize our effort. Indeed, if we consider the worst-case scenario of a strategy, where APTs are slightly faster to exploit than enterprises to patch (i.e. on the same month), the probability of being compromised significantly overlaps with the more relaxed strategy of waiting for a CVE to be published.

All-speedful defenders 
are as unrealistic as all-powerful 
attackers, thus it seems perfectly 
rational to save time and money and only 
rush to update for 
\emph{disclosed} vulnerabilities and before that 
do nothing. That is what most company seems 
to do and get lambasted for.

\emph{What is the answer?} If the "keep your systems up to date" best 
practice is only effective when the 
enterprises spend most of their resources 
on fixing and maintaining their suppliers' 
software instead of servicing their own 
customers, then why do we stick with this 
recommendation? 

This is a best 
practice that can be included in the 
"security 
theatre"~\cite{schneier2009beyond}, 
measures that provide minimal benefits and 
are not worth the cost. 

As Kamp points out in the mentioned CACM 
columns~\cite{kamp2021software} this 
industry `best practice' allows cyber-insurance companies to reject requests for 
coverage. Most importantly, it allows software companies to shift 
liability towards their "lazy" users.
For the software vendors (starting from the 
big four), this best practice is the most
desirable 
because releasing an update that "improves" 
an unknown feature of a product is much 
easier than implementing it as a secure 
application with mechanisms like automatic network segmentation and escalation and execution confinement~\cite{massacci2021solarwinds} and without unnecessary frills. 

We agree with Poul-Henning Kamp, 
it is a jolly good time 
to introduce the `right to not update' \cite{massacci2021solarwinds} and return
the liability of defective software 
building to the software builders and not 
to the software users \cite{kamp2021software}. 

Dwellers of the U.S. Capitol or Brussel's Berlaymont should take  
notice of the copy of the old Code of 
Hammurabi reported in ~\Cref{tab:hammurabi} along with its counterpart in the IT reign.
It is for the future to align its liability laws to the past.

\begin{table}
\caption{The IT Code vs the Building Code of Hammurabi}
\begin{tabular}{p{0.25\columnwidth}p{0.35\columnwidth}p{0.3\columnwidth}}
\toprule
What Happens & Code of Hammurabi (ca. 1780 BC) & Code of IT Industry (2022 AD) \\
\midrule
If a [builder] did not construct properly the house for a fellow [\ldots] and a wall fell & 
that builder 
shall make that wall sound using his own 
silver & that builder 
will suggest another house he built and that fellow should pay to 
move one's goods to the other house. \\
If the house fell and it ruins goods & [the builder] shall make compensation for all that has been ruined & \emph{unthinkable} \\
\\
if the builder wants change the walls or close a window of the house after the sale & \emph{unthinkable} & the fellow shall adapt its living to the new house at one's own expenses \\
\bottomrule
\end{tabular}
\label{tab:hammurabi}
\end{table}

To summarize: If ``how to have more security?'' is the question, software updates are not the answer.

\bibliographystyle{ACM-Reference-Format}
\bibliography{ms}

\end{document}